\documentclass[sigplan, 10pt]{acmart}
\AtBeginDocument{%
  \providecommand\BibTeX{{%
    \normalfont B\kern-0.5em{\scshape i\kern-0.25em b}\kern-0.8em\TeX}}}
    
\setcopyright{acmcopyright}
\copyrightyear{2021}
\acmYear{2021}

\usepackage{graphicx}
\usepackage{color}
\usepackage{balance}
\usepackage{mathtools}
\usepackage{balance}
\usepackage{listings}
\usepackage[linesnumbered,ruled]{algorithm2e}
\usepackage{bm}
\usepackage{url}
\usepackage{paralist}
\usepackage[normalem]{ulem}
\usepackage{microtype}
\usepackage{titlesec}

\makeatletter
\renewcommand\section{\@startsection{section}{1}{\z@}%
                       {-8\p@ \@plus -4\p@ \@minus -4\p@}%
                       {6\p@ \@plus 4\p@ \@minus 4\p@}%
                       {\normalfont\large\bfseries\boldmath
                        \rightskip=\z@ \@plus 8em\pretolerance=10000 }}
\renewcommand\subsection{\@startsection{subsection}{2}{\z@}%
                       {-8\p@ \@plus -4\p@ \@minus -4\p@}%
                       {6\p@ \@plus 4\p@ \@minus 4\p@}%
                       {\normalfont\normalsize\bfseries\boldmath
                        \rightskip=\z@ \@plus 8em\pretolerance=10000 }}
\renewcommand\subsubsection{\@startsection{subsubsection}{3}{\z@}%
                       {-4\p@ \@plus -4\p@ \@minus -4\p@}%
                       {-1.5em \@plus -0.22em \@minus -0.1em}%
                       {\normalfont\normalsize\bfseries\boldmath}}
\makeatother

\newcommand{\para}[1]{\noindent\textbf{#1}.}

\newcounter{inlineenum}
\renewcommand{\theinlineenum}{\alph{inlineenum}}

\titlespacing*{\section}{0pt}{1.0ex plus 0.5ex minus .2ex}{1.0ex plus .1ex}
\titlespacing*{\subsection}{0pt}{1.0ex plus 0.5ex minus .2ex}{1.0ex plus .1ex}
\begin{document}
%
%\title{SoK: Uses of Trusted Execution Environments in Blockchain Based Financial Applications}
\title{Lessons Learned from Blockchain Applications of Trusted Execution Environments and Implications for Future Research}
\begin{abstract}

Modern computer systems tend to rely on large trusted computing bases (TCBs) for operations. To address the TCB bloating problem, hardware vendors have developed mechanisms to enable or facilitate the creation of a trusted execution environment (TEE) in which critical software applications can execute securely in an isolated environment. Even under the circumstance that a host OS is compromised by an adversary, key security properties such as confidentiality and integrity of the software inside the TEEs can be guaranteed. The promise of integrity and security has driven developers to adopt it for use cases involving access control, PKS, IoT among other things.
 Among these applications include blockchain-related use cases. 
The usage of the TEEs doesn't come without its own implementation challenges and potential pitfalls.
In this paper, we examine the assumptions, security models, and operational environments of the proposed TEE use cases of blockchain-based applications. The exercise and analysis help the hardware TEE research community to identify some open challenges and opportunities for research and rethink the design of hardware TEEs in general.

\end{abstract}

%  \and Crypto-currency

% options: applications, systems, infrastructures
% %\titlerunning{TEE in Crypto-economic Applications}
% \titlerunning{TEEs in Blockchain based Financial Systems}
% \title{Trusted Execution Environment in Blockchain Based Computing: a taxonomy,
% review and future directions}
% \title{Trusted Execution Environment in Defi: a taxonomy, review and future directions}
%
% \title {SoK: Use case of blockchain in secured environments}
% \title {SoK: Use case of Trusted Execution Environment in Blockchain}
% \title {SoK: Uses of Trusted Execution Environment for Defi}


\author{Rabimba Karanjai}
\affiliation{%
   \institution{University Of Houston}
   \state{TX}
   \country{USA}}
\email{rkaranjai@uh.edu}

\author{Lei Xu}
\affiliation{%
  \institution{University of Texas Rio Grande Valley}
  \state{Texas}
  \country{USA}
}
\email{xuleimath@gmail.com}

\author{Lin Chen}
\affiliation{%
  \institution{Texas Tech University}
  \state{Texas}
  \country{USA}
}
\email{Lin.Chen@ttu.edu}

\author{Fengwei Zhang}
\affiliation{%
   \institution{Southern University of Science and Technology (SUSTech)}
   \state{Shenzhen}
   \country{China}
}
\email{zhangfw@sustech.edu.cn}

\author{Zhimin Gao}
\affiliation{%
  \institution{Auburn University at Montgomery}
  \state{Alabama}
  \country{USA}}
\email{mtion@msn.com}

\author{Weidong Shi}
\affiliation{%
   \institution{University Of Houston}
   \state{TX}
   \country{USA}}
\email{wshi3@uh.edu}

\maketitle              % typeset the header of the contribution

\section{Introduction}\label{sec-introduction}
Modern computer systems tend to rely on large trusted computing bases (TCBs) for operations. Typically, a TCB includes complex software components such as the OS kernel, device drivers, privileged services, and libraries. Systems built on top of a bloated TCB tend to be more vulnerable to exploits and have large attack surface compared with the systems with small-sized TCBs. Such systems including the application software sitting on top of the platforms are more likely to be exposed to attacks, opening more doors for potential security deficiencies and effective exploitation. 

To address such TCB bloating problems, hardware vendors have developed mechanisms to enable or facilitate the creation of a trusted execution environment (TEE) in which critical software applications can execute securely in an isolated environment. Even under the circumstance that a host OS is compromised by an adversary, key security properties such as confidentiality and integrity of the software inside the TEEs can be guaranteed. 

The security properties of a TEE implemented as hardware features protect the confidentiality and integrity of computations that take place inside it. 
In addition, to enforce isolated execution, a TEE abstraction often defines support for basic security operations such as remote attestation for verifying that a piece of software is executed by a genuine CPU using a TEE, secure provisioning of code and data (including cryptographic keys) into the TEE, and trusted communication channels for retrieving computing results and outputs from the TEE. These security properties make the hardware TEEs suited for a wide range of applications. 

Among these applications include blockchain-related use cases. Blockchain-based applications of the TEEs are somewhat unique because of the operational environments (often decentralized and open to the public as a computing infrastructure) and security requirements. There are clear financial motivations for adversaries to attack TEE based blockchain environments. Leveraging the security properties like isolation and attestation, one can construct trusted nodes for blockchain based applications. Recognizing the TEE's potential,  researchers have proposed to apply the TEEs to support various blockchain components and functionalities including distributed consensus protocols, smart contract execution, exchange services, payment network, oracle service provider, cryptocurrency wallet, etc. For instance, distributed consensus protocol design can leverage the TEEs to improve performance~\cite{poet}, and one can build reliable oracle services to introduce external information into a blockchain~\cite{10.1145/2976749.2978326}. 

In this paper, we analyze the blockchain use cases of the TEEs and identify the security models of these use cases. The goal is to take the lessons learned from these use cases and systems, and rethink the TEE designs, security requirements for the TEEs,  and targeted operational environments. Although we focus on blockchain applications in this paper, the same exercise can be applied to other blockchain applications. The contributions of this work include:
\begin{inparaenum}[(i)]
    \item We summarized some of the TEE use cases in the construction of blockchain-based systems and applications, and analyzed their security models;
    \item We discussed the challenges of utilizing the TEEs in these applications; and
    \item We identified future TEE research directions and opportunities to enhance the hardware TEEs to meet the security and operational needs of applications.
\end{inparaenum}

\section{Commercial Hardware TEE platforms}\label{sec-background}
In this section, we briefly describe the popular commercial TEE platforms. Readers who are knowledgeable of these platforms can skip this section.

\subsection{Intel SGX}

Intel Software Guard Extensions (SGX)~\cite{10.1145/2487726.2488368,intelsgx} is the instruction set architecture that enables the creation of an enclave, a TEE. %A SGX enclave ensures the integrity and the confidentiality of the code and the data of the program running inside SGX enclave. 
SGX enclaves are isolated memory regions of code and data residing in a protected memory space called EPC (enclave page cache). Only the code within the enclave can access the sensitive content stored inside the protected memory regions. All external accesses will be denied. 
%SGX supports multiple enclaves over a single platform. 

For managing the protected memory regions, EPC is split into pages of equal size. These pages can be assigned to different enclaves using the SGX instructions. 
SGX utilizes a hardware memory encryption engine to protect the EPC memory so that all the EPC pages stored in the main memory are encrypted. Memory access control is enforced at the hardware level where all the read and write requests to the EPC pages are checked for violation of access management and handled by the memory encryption engine. All non-enclave accesses are prohibited. 
The enclave entry point is predefined in a compilation phase. 
SGX can detect violations of memory integrity or replayed memory content. However,  SGX does not hide the locations where memory updates and reads refer to, which could be exploited as a side-channel.  SGX supports remote attestation~\cite{intelapi} that allows establishing a secure communication channel between a secure enclave and a client. Remote attestation is achieved through a centrally managed Intel Attestation Service (IAS) that verifies the enclave and provides the client the assurance of enclaves’ authenticity. This means to use SGX enclave, one must trust the Intel Attestation Service. To protect privacy, Intel uses a group signature-based scheme, called EPID~\cite{epid2016} for authentication. %SGX enclaves run in user mode and are protected from a potentially hostile or compromised OS. It reduces the attack surface and offers protection against attackers who have gained privileged access to a platform.
SGX2~\cite{10.1145/2948618.2954330} extends the SGX instruction set to enable dynamically managed memory within an SGX enclave. It allows developers to use new SGX2 features for implementing functions such as dynamic heap management, stack expansion, and thread context creation.

SGX DCAP~\cite{Scarlata2018SupportingTP} is an extension to SGX remote attestation support. It allows third-party users of SGX to manage their own attestation infrastructure. This new feature is designed for data center operators and cloud service providers.
It can potentially benefit application models working in a distributed environment. % such as peer-to-peer networks by not relying on a centralized attestation scheme like the Intel remote attestation services. 
Currently, DCAP is only supported by the Intel Xeon E Processors.  

\subsection{AMD SEV}
Recognizing the need for a secure cloud computing environment, AMD has developed its own TEE support, named Secure Encrypted Virtualization (SEV)~\cite{208506,amd2016}. AMD SEV integrates hardware-enabled memory encryption capabilities with the existing AMD-V virtualization architecture to support encrypted virtual machines. Encrypting virtual machines
can help protect them not only from certain physical attack threats (e.g., physical memory probing) but also from other virtual machines or even the
hypervisor itself. It specifically targets customers of data centers and cloud computing environments. 

SEV depends on AMD Secure Processor (AMD-SP) for key management and remote attestation. AMD-SP is an ARM-based processor (Cortex-A5) integrated with the AMD SoC as a dedicated security subsystem. 
Main memory encryption is performed via dedicated hardware in the on-die memory controllers called Secure Memory Encryption (SME). SME is a general-purpose main memory encryption mechanism integrated into the AMD CPU architecture, and it is designed to provide high performance for server workloads. % (with SME). 

% SME supports either a full memory encryption model where all of DRAM is encrypted or a partial memory encryption model where only selected regions are encrypted such as memory spaces corresponding to the guest virtual machines.  To protect SEV enabled guests, AMD-SP firmware assists in the enforcement of security properties such as the authenticity of the platform, attestation of a launched guest, and confidentiality protection of the guest’s data. 

Platform authentication prevents adversaries from masquerading as a legitimate AMD SEV enabled platform. The authenticity of the platform is proven with its platform identity key signed by AMD. Similar to the SGX attestation service, AMD provides its own centralized service for remote attestation~\cite{amdcek}.  %Using a signature generated by the firmware, attestation of the guest launch proves to the guest owners that their guests are securely launched with SEV enabled.  After attestation, a guest owner can ensure that the host does not interfere with the initialization of SEV enabled guest. Then the guest owner can transmit confidential information such as data decryption keys to the guest. 
%The confidentiality of the guest is accomplished by encrypting memory with a memory encryption key that only the SEV firmware knows. 

SEV-ES (encrypted state) offers an extended feature to protect CPU register states. During context switch (transition between a hypervisor and a guest virtual machine), the virtual machine register state will be encrypted to protect the confidentiality of the data in the guest memory. The recent extension, called SEV-SNP (securely nested paging) adds new hardware-based security protection features on top of SEV and SEV-ES~\cite{amdsnp}. SEV-SNP enables memory integrity protection to prevent attacks such as replay of encrypted data, data corruption, memory re-mapping, memory aliasing, TCB rollback, etc. 
Furthermore, SEV-SNP implements additional security features to protect interrupt behavior, and reduce vulnerability against certain side-channel attacks like the poison of BTB (branch target buffer) from hypervisor. A comparison between SGX and SEV can be found in~\cite{10.1145/3214292.3214301}.

\subsection{ARM TrustZone}
TrustZone is a security extension to the ARM architecture with modifications to the processor, memory, and I/O devices~\cite{armtrustzone}.  The most important architectural change at the processor level consists in the introduction of two protection domains designated by the name of: the secure world and the normal world. The memory system has been extended with TrustZone security features by adding hardware components called,  TrustZone Address Space Controller (TZASC) and TrustZone Memory Adapter (TZMA). The TZASC can be used to configure specific memory regions as secure or non-secure. It partitions the main memory into different regions and manages its association with a specific world (normal or secure). 

% For instance,   applications running in the secure world can access memory regions associated with the normal world, but not the other way around.  The TZMA provides a similar function but for off-chip SRAM or ROM. 

The TrustZone-aware Memory Management Unit (MMU) allows for each world to have its own virtual-to-physical memory address translation tables. Memory isolation is also supported at the cache level with each cache line tagged to indicate under which world that cache line access has been performed.  

Due to the ARM's license model, the adoption of TrustZone depends on specific SoC implementation. Xilinx Zynq and NXP i.MX6 are examples of SoCs that provide full support for the TrustZone technology. TrustZone is an enabling technology or building block for creating ARM-based TEEs. TrustZone by itself does not implement TEE attestation. TEE implementation built on top of TrustZone can support attestation (e.g.,~\cite{DBLP:journals/scn/WangZY20}), and other security features like secure communication channel.	
Examples of ARM based TEEs are OP-TEE~\cite{optee} and Open-TEE~\cite{McGillion_2015}.

\subsection{FPGA based TEE}
FPGA (field programming gate array) provides an alternative general-purpose platform for building TEEs, which leverages the reconfigurability of the FPGA technology
to support security guarantees such as confidentiality and integrity. 
In FPGA-based design, secure enclaves can be implemented in the programmable logic as physically isolated domains. Physical isolation is a major advantage in terms of security. %Physical isolation can prevent most side-channel attacks that plague the other platforms. 
Programmable logic can function as a TEE as long as security features such as attestation, memory encryption, and proper access control mechanism of memory accesses are implemented. All of them can be implemented as programmable logic modules. Typically, FPGA itself does not implement TEE attestation. PUF based attestation can be enabled by specific FPGA TEE design~\cite{Patten2018AFF,6881436}. FPGA allows general-purpose computing through the integration of a soft-core CPU.  %There are many soft-core options (open source and proprietary) including but not limited to Cortex-M1, MicroBlaze, SPARC V8, RISC-V, and etc. 

\subsection{Summary of commercial TEE platforms}

%\tablename~\ref{tbl-tee-cmp} compares different TEE platforms. 
In general, certain observations can be made regarding commercial TEEs. 

{\bf Firstly}, remote attestations often depend on centrally managed services. 
{\bf Secondly}, each platform appears to target different application sectors, for instance, ARM for mobile devices (limited presence of ARM devices in the server market), AMD-SEV for data centers and cloud. Intel also has started to focus on the cloud market by introducing DCAP. %Development of the kernel source code for SGX seems to indicate that continued support of EPID attestation in future Intel products is uncertain. 
{\bf Thirdly}, the boundary between hardware support for the TEEs and software security features has not been finalized. The evolving security features of SEV, SEV-ES, and SEV-SNP provide an example. Often security primitives in the TEE hardware do not protect against side-channel attacks (e.g., memory access patterns), which the hardware vendors consider responsibility for software developers~\cite{intelside}. %Disclosed attacks in recent years make it a continuous dialogue between the hardware TEE designers and the software developers on the boundary of TEE hardware and software, in particular protection against architectural side-channels.   
{\bf Fourthly}, FPGA offers a unique solution for TEEs because it can physically isolate the TEEs from the untrusted components. 
{\bf Fifthly}, from the portability aspect, AMD TEEs provide the best environment because existing software can be supported out-of-box. FPGA may be the least portal option because of its unique development environment.

\section{Use Case Analysis}

In this section, we provide a summary on some use cases of the TEEs in decentralized ledgers and blockchains. 
The goal is to understand how these applications apply the TEEs, the key security assumptions behind them, and the TEE functionality requirements crucial for the application use cases. It is by no means intended as a survey of TEE applications in blockchains. 
%These use cases span a range of blockchain components and functionalities including distributed consensus protocols, smart contract execution, exchange services, payment network, oracle, wallet, etc. 

Several research efforts aim to support efficient BFT protocols by leveraging the SGX enclaves (e.g., ~\cite{8416471,10.1145/3064176.3064213}). 
The security primitives provided by the enclaves can reduce the number of replicas and/or communication phases for running BFT protocols. To minimize TCB size, in FastBFT, only cryptographic keys and secret-shares are protected by the enclaves using SGX's crypto library.

PoET ~\cite{poet} is a Nakamoto-style consensus algorithm ~\cite{nakamoto2008bitcoin}. It replaces the PoW by a wait time randomly generated by a SGX enclave. This elapsed time consensus assumes that the TEE has a trusted timer. To handle potential threats posed by compromised SGX enclaves, the PoET protocol uses a rate limit function based on z-test to restrict the number of blocks that a participant can publish in any particular larger set of blocks. Another SGX for distributed consensus example is Proof of Luck ~\cite{milutinovic2016proof}. In Proof-of-Luck (PoL) design, participants use the SGX enclaves to perform routines to maintain a blockchain. PoL assumes an economic security model such that compromising many enclaves is prohibitively expensive. It contains a defense mechanism against scenarios that an adversary compromises a limited number of  enclaves. 

In ~\cite{bowman2018private}, the authors describe private data objects and apply the SGX enclave to protect data confidentiality during smart contract execution. It enables mutually untrusted parties to run smart contracts over private data (PDOs). SGX enclaves are used to preserve data confidentiality, execution integrity and enforce data access policies. A distributed ledger verifies and records the transactions produced by the PDOs. This way, the scheme enables stateful functionalities for otherwise stateless SGX  enclaves. 

Obscuro ~\cite{cryptoeprint:2017:974} is a Bitcoin mixer that utilizes the SGX enclaves. Leveraging the TEE’s confidentiality and integrity guarantees, Obscuro aims to ensure the correct mixing operations and the protection of sensitive information like private keys and mixing logs. For attestation, Obscuro assumes that the measurement report is distributed to a public bulletin board (can be provided by a distributed ledger). The security model assumes that an adversary can launch an attack that rewinds enclave state. To mitigate such an attack,  Obscuro stores no permanent states. To defend against potential side-channel exploits, the authors mention three approaches, (i) performing address space layout randomization (ASLR) protection inside the enclave using SGX-Shield; (ii) using non-vulnerable cryptography libraries resistant to cache side-channel attacks; and (iii) applying Zigzagger compiler to reduce control flow footprint.

TEEs are ideally suited for protecting cryptocurrency wallets. 
SBLWT	~\cite{8412192} is a	secure and lightweight wallet based on Trustzone for cryptocurrencies.  It is more portable compared with the hardware wallet, and more secure than the software wallet. Leveraging TrustZone isolation, SBLWT protects the private key and the wallet’s address from being compromised by adversaries no matter whether the host environment including the main OS is malicious or not. 
TEEOD enables FPGA based TEEs ~\cite{pereira2021trusted}. TEEOD implements secure enclaves in the programmable logic (PL) by instantiating a customized and dedicated security processor per application use case on demand. It can instantiate up to six simultaneous enclaves. Each enclave includes a softcore CPU and uses shared memory for communications. A Bitcoin wallet implementation is demonstrated using TEEOD.

These applications make a range of assumptions. Below, we provide a summary of common security assumptions and/or security requirements regarding the TEEs. 
\begin{inparaenum}[(i)]
    \item It is assumed that the adversary can control everything outside the TEEs including an interface to the TEEs. Further, an adversary can manipulate input and world view to the TEEs. For the FPGA-based TEEs, an adversary can launch a bitstream re-programming attack. Almost all the TEE use cases have considered such attack scenarios. 
    \item Certain use cases require that the TEE has a trusted real-time clock or equivalent functionality (e.g.,~\cite{10.1145/2976749.2978326,poet}).  An additional requirement is that TEE is equipped with a monotonic counter which can never be decreased in value, even after the TEE is reset~\cite{poet}. 
    \item Certain use cases assume that TEE can provide a means to prevent undetected replay attacks or rewinding attacks. In the case of a replay attack, an adversary saves the encrypted state of a trusted component and starts a new instance using the exact same state to reset the component. The SGX hybrid BFT design requires such protection~\cite{10.1145/3064176.3064213}.  Obscuro mixer design assumes that an adversary can launch an attack by rewinding enclave state~\cite{cryptoeprint:2017:974}. A mitigation strategy to such attacks is to consider whether the design can be stateless, which is a strategy used by some use cases (e.g.,~\cite{cryptoeprint:2017:974}). 
    \item A common assumption is that an adversary cannot mount a scalable attack to the TEEs (e.g.,~\cite{bowman2018private,203890}). Under this assumption, an adversary can only compromise a limited number of TEEs and gain full control of these machines including compromising the TEE's signing and attestation keys. This means that attackers can masquerade as the compromised TEEs and issue valid attestations using their identities. However, it is assumed that the attackers cannot arbitrarily forge enclave identities. A question is that \textit{is it possible multiple adversaries compromise a large percentage of TEEs in a single blockchain system even individual adversary can only subvert a limited number of TEEs?} %  SGX EPID signatures are linkable (permits anyone to determine whether two signatures were generated by the same CPU). 
    \item All existing use cases assume that the TEE hardware vendors are trusted, which includes the centrally managed remote attestation services. 
\end{inparaenum}

\section{TEE Attack Surface}\label{sec-tee-security-risks}

TEEs are vulnerable to certain attacks. %In this section, we provide an overall picture of recently reported attacks to the TEEs with a focus on SGX but also covering attacks to TrustZone and AMD SEV technologies. 
Often, it is assumed that attackers to the TEEs include both local attackers who control the hardware and software platform for hosting enclaves and remote attackers who successfully compromise a victim platform used for running enclaves and consequently gain full privileged control of the victim. In general, it is assumed that the TEE hardware vendors are trusted. At least, it may be safe to assume that the TEE manufacturers would not launch all-out indiscriminate attacks that target their TEEs. However, the risk from the TEE hardware vendors cannot be completely ignored which we will examine in the next section. 

For both local and remote attackers, one can assume that the attackers can intercept the input to and the output from the TEEs. Attackers can manipulate, modify, insert, delete data communicated with the TEEs, or change the ordering of data. Furthermore, attackers can halt or reset the physical machines at any time. In case that a security model considers physical attacks, the attackers can intercept DDR bus transactions or manipulate  DDR bus. Through I/O devices, attackers can issue rogue DMA transactions or hardware signals to tamper with the TEEs.  Most application security models do not consider physical attacks because of the resources and skills required by such attacks. 

We consider that TEE security goals include, confidentiality, integrity, attestation, protection against DoS, and privacy. Privacy is included as an independent goal to cover certain information leakage scenarios like fingerprinting the application executed by a TEE. 
In the literature, attacks are often reported with a focus on a specific TEE platform, for instance, SGX, TrustZone, or SEV. However, many side-channels exploits are inherent risks universal to modern high-performance micro-architecture designs thus affect almost all the TEE products (e.g., cache timing-based attacks ~\cite{10.1109/SP.2015.43, 191010, 10.1007/11605805_1, 184415}, out-of-order speculative execution and other micro-architectural features  ~\cite{Evtyushkin2018BranchScopeAN,8806740,Huo2020BluethunderA2,220586}).   These exploits can not only compromise both confidentiality and integrity protection of the TEEs but also attestation signing keys embedded in a TEE platform. Some recent surveys of TEE attacks and countermeasures can be found in ~\cite{Nilsson2020ASO,10.1145/3456631}.

\section{Challenges from the Use Cases}
\label{sec-challenges}
Analysis reveals certain shortcomings or uncharted areas in TEE designs and research from the use case aspects. 

\subsection{Reliance on centralized infrastructure and single trusted party}
Most TEE solutions rely on the TEE hardware vendors' centralized services such as remote attestation.
%in particular services related to TEE remote attestation. 
In decentralized systems and applications, these services are centralized building blocks that can make the entire system exposed to certain risks such as potential single point of failure and tampering from the hardware TEE vendors. One can assume that the hardware vendors will not engage in activities that undermine the security of their own products. However, in the unlikely event, vendors can be compelled to disclose TEE identities or perform DoS targeting specific TEEs or TEE applications. Such concern cannot be completely ignored. According to ~\cite{Swami2017IntelSR}, weakness in the EPID implementation could make it possible to target specific TEEs or TEEs used in a specific application. If TEEs are hosted by a service provider like the cloud, the risk must be higher because it is not that difficult for a service provider to fingerprint a TEE application using side-channel information, for instance, network ports and traffic patterns, performance counters. Then DoS can be launched targeting specific TEE users or specific TEE based systems. 

SGX DCAP has the potential to facilitate more flexible enclave attestation schemes, a step forward in reducing dependency on centralized attestation infrastructure. At this moment, it primarily focuses on the cloud and data centers. Whether it is well suited for P2P applications or decentralized systems remains to be demonstrated. 

\subsection{Missing economic based TEE security model}

\begin{figure}
    \centering
    \includegraphics[width=3in,height=1.2in]{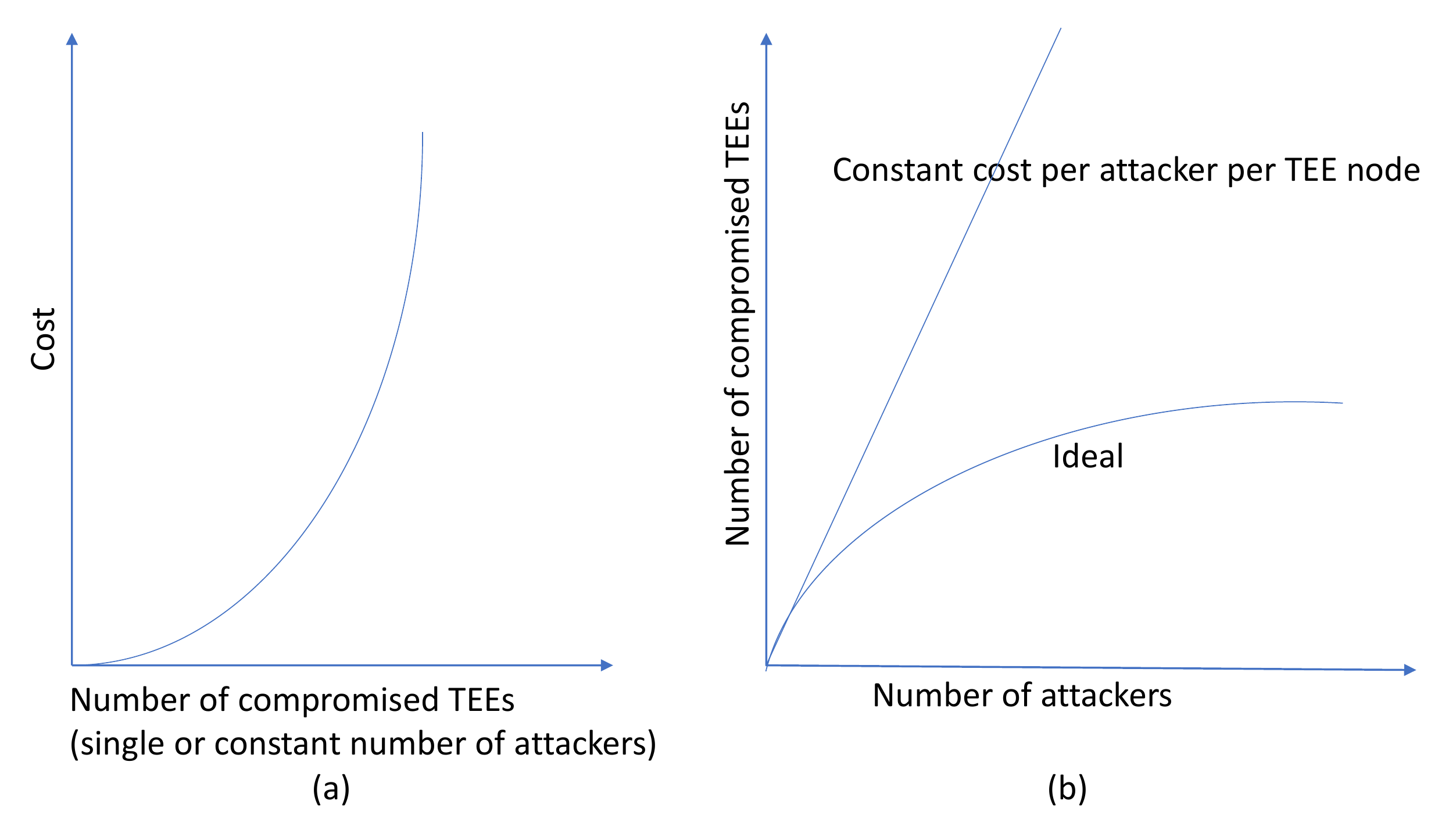}
    \caption{Economic security model of hardware TEEs: (a) single or constant number of adversaries; (b) multiple adversaries in a P2P or decentralized system.}
    \label{fig:model}
\end{figure}

Many use cases assume that there exists an economic-based TEE security model where financial cost grows in such a pace that it prevents adversaries from achieving large-scale breaches of TEE security.

% Rarely, this has been a requirement for TEE design and development because such economy driven security model for the hardware TEEs has not been well understood and studied. 

One could assume that for a single attacker if an exploit requires significant investment in physical resources and sophisticated skills like an attack on the DDR bus, it is reasonable to assume that only a very limited number of TEEs could be compromised using such an attack. However, most documented attacks in the literature to the SGX enclaves and AMD SEV are software-based attacks, which can be launched by local attackers. So if the attackers are sufficiently motivated by financial gains (for instance, using TEEs in Proof-of-Work blockchain consensus to compute winners of block reward), a large scale attack scenario launched by many individual attackers is plausible in the real-world, even for unskilled attackers if they can download attack software and run it to compromise local TEEs that they have direct control in a P2P or decentralized system.

In summary, it is elusive to develop a quantitative economy based TEE security model. Very little research has been done. Incorporating such a model as requirements into hardware TEE design is equally challenging because of the vague assumptions and uncertainty in quantifying security properties using economic terms. 

\subsection{Unclear security boundary between hardware TEE and software}
Similar to any computer architecture research on hardware-software interfaces, security research, and applications based on the commercial hardware TEE platforms once again illustrate the challenges in drawing a boundary in TEE security between the software and hardware. What security protections and which kinds of security primitives should be ideally implemented by a hardware TEE? What security features should primarily be software developers' responsibilities? A clear understanding of these questions is crucial not only for designing future hardware TEEs but also for developing a general hardware-software co-design framework for the TEEs. Arguments and counter-arguments often can be found from both sides involving hardware TEE designers and architects, security researchers, and application developers. If the history of computer architecture can be applied as a prediction of the future, there will never be a clearly delineated boundary. The security boundary will remain open and dynamic, sometimes heatedly debated.  However, nothing will prevent researchers and developers to explore the hardware TEE design space and experiment with different options. As we have described earlier, blockchain-based applications are unique in contributing to this discussion because of their unique security requirements and decentralized operational environments. 

\subsection{Lack support of rapid detection and response to compromised TEEs}
Many use cases and applications of the TEEs assume that compromised TEEs can be quickly detected and swiftly removed from the system. Despite a large body of published work, there is a lack of research in efficient detection of compromised TEEs and rapid response mechanisms. Likely this lack of research is partly due to the closed and centralized remote attestation services where only the vendors can remove the compromised TEEs from attestation. When the privacy of TEE identities is a requirement,  the challenge will be more significant because the requirements on protection of TEE privacy and rapid detection/response to compromised TEEs may be conflicting goals in designing the TEE identity and attestation schemes. Linkable TEE signatures will facilitate rapid response to compromised TEEs. However, it may raise privacy concerns. A zero-knowledge proof based hardware TEE identity scheme can provide full privacy protection. On the other hand, it makes detection and removal of the compromised TEEs from a distributed system a much difficult task. In addition, how to prevent such detection and response mechanisms from being abused by malicious attackers, who can leverage the existence of such services for denial of service attacks. In case TEE attestation service is decentralized, one may face additional challenges such as reaching consensus on rogue TEE detection and agreements in responses. Overall, it is not entirely clear how a hardware TEE platform can balance these requirements like privacy, the operational environment, etc to realize a detection and response scheme for compromised or rogue TEEs that is rapid, privacy-preserving, flexible enough to meet the diverse operational needs based on the applications.  

\subsection{Lack of standards and inter-operability}
%Lack of standards and inter-operability support across TEEs of different hardware vendors. 
A great challenge that hinders hardware TEEs from gaining broader adoption than what has been achieved so far is that there are no well-established and well-accepted standards that the vendors agree to follow. Commercial TEEs such as SGX, SEV, and TrustZone are all based on proprietary designs and implementations. This creates significant barriers to supporting interoperability across the TEEs. The research community has realized the advantages of a heterogeneous TEE environment to improve infrastructure resilience and security posture against attacks. However, many challenges remain for developing applications over a heterogeneous TEE environment. 

% Developers may face issues like incompatible attestation schemes, heterogeneous root-of-trust, lack of formal semantic of security primitives offered by different hardware TEE vendors, etc. Moreover, micro-architecture designs of TEEs are closed from the security community as trade secrets and protected intellectual properties. This makes the task of assessing security risks and threat modeling of different hardware TEEs very difficult. 

\section{Opportunities for Enhancement and Future Research }
\label{sec-future}

Despite the challenges, focused studies of application use cases, their requirements, security assumptions as well as hurdles behind them provide insights of potential topics for future research of TEE designs and implementations. These improvements if adopted could benefit a broad range of applications beyond what we have used for analysis. 

\subsection{New hardware TEE security primitives}
There are opportunities to develop new security primitives and/or enhance the existing security primitives provided by a hardware TEE platform. Here we provide some examples based on the use case studies. 

\para{Trusted and accurate wall time clock} There is a clearly identified need for a secure TEE based wall time clock. Many factors could affect a computer's wall clock time. Variations of these factors can cause severe drift of clock time, sometimes even lead to the problem with NTP synchronization.  Some of these factors depend on the motherboard chipset. In the past, control for variable speed of processors, front-side bus spread spectrum oscillators, mis-detection in TSC all could contribute to inaccurate clock time and large drift in normal situation even without an adversary. An accurate and tamper-resistant TEE wall time clock is desired by many applications (for instance, Google's noSQL database applies GPS time for data synchronization). To reduce reliance on NTP, which can be manipulated by attackers, even chip size atomic time source could be integrated with a hardware TEE platform to provide accurate and secure TEE time. 

%\subsection{Extension of TEE to support continuous state}
\para{Extension of the hardware TEE to support continuous state} Another security primitive is to support non-rewindable and persistent TEE storage space for storing application states that can not be reverted. One of the uses cases of hardware backed secure storage and what it can store and utilize has been on the topic of persistent data integrity. However a lot of the security mechanisms hold up when the system is in an uninterrupted powered state. In real life however power crashes,system reboot are pretty common, making these assumptions unrealistic. State-protected modules must securely store their state. 
In the topic of state continuity, Intel SGX and other hardware-based TEE's are prone to state module replay attacks. This assumes that the attacker has complete control over the untrusted software stack and can replay the enclaves by manipulating execution control. Strackx et al in Ariadne \cite{197191} proposes a system that achieves state continuity and gives a solution to this problem by providing a minimal attack surface. This is done by ensuring the following three properties of rollback prevention, liveliness detection of the system and also ensuring an execution is not interrupted before it reaches end of cycle. However, this subject has not received much attention, showing mostly Ariadne \cite{197191} and Parno et al. \cite{parno2011memoir} are one of the few who has proposed solutions. %We believe this field deserves more research to come up with solutions that does not rely on continuous machine state.
Because SGX enclave is stateless, many blockchain applications with such needs are forced to customize the designs to be stateless, which is less ideal and possibly may introduce new attack surface due to extra design complexity. 

% Other applications may benefit as well from such security feature. As discussed above, alternative approaches are to, integrate other secure hardware component like a TPM together to achieve  non-rewindable states for the TEEs or leverage the immutability feature of blockchains. However, these alternative approaches are far less than build-in support of  non-rewindable states.  

\subsection{Decentralized TEE identity management and attestation}

There has been an explosion of research in decentralized and self-governed identity management applying the features of blockchains. In contrast, not much has been done to extend the concept to the TEE identity.  Identity management and attestation for hardware TEE platforms remain centralized and rely on the vendors' centrally managed attestation services. Existing research on remote attestation using commercial TEE platforms mainly focus developing API wrappers or proxy attestation service that can hide the heterogeneity  of hardware TEE attestation. 
An example is ARM research's Veracruz project ~\cite{veracruz} that aims to support cloud-based IoT applications with heterogeneous TEEs. The project, under Confidential Computing Consortium~\cite{armccc}, demonstrates some of the hurdles that one has to overcome in order to support an infrastructure of heterogeneous TEEs.

% for instance, how to manage diverse attestation schemes and compatibility issues in heterogeneous root-of-trust. 
To conclude, how to enable a truly decentralized infrastructure for hardware TEE identity management and attestation looks like a promising direction of research, specifically innovative approaches that can balance the needs of privacy, decentralization, detection and rapid response to compromised hardware TEEs. 

\subsection{Programmable TEE security}

Various techniques such as address space randomization (both code and data), hiding of control flow patterns using compiler, software based ORAM, side-channel resistant cryptographic libraries are proposed as defense on attacks to the SGX enclaves (e.g., ~\cite{SeoLKSSHK17,DBLP:journals/corr/abs-1709-09917,DBLP:conf/ndss/SasyGF18,203698}). Performance overhead to the TEE applications after applying some countermeasure techniques varies from less than 10\% to potentially hundred of times slow down (see ~\cite{10.1145/3456631} for a survey of countermeasures and performance overhead). In most cases, TEE applications have to be executed in single thread mode.  To support efficient execution of diverse applications with different security requirements, future TEEs can make the hardware TEE platform more configurable or programmable. 

A hardware TEE platform can provide a programmable security engine to the developers to implement advanced security features like address space randomization or ORAM (Oblivious RAM)  in hardware. 

% For instance, a programmable memory controller can be used to randomize memory operations, which can significantly increase the performance of the TEE applications that prefer to use address space randomization and ORAM as a defense. 

This can be achieved using a FPGA integrated memory controller for the TEEs. Based on the need, TEE application developers can supply a FPGA bitstream that implements certain enhanced security protection mechanism like a lightweight and customized ORAM layer. The measurement report will be extended to include configuration settings and digital signatures of the security programs applied to such memory controller.  This framework can provide a more flexible security environment with high performance to meet the diverse security and performance needs of TEE applications. 

\subsection{Hardware TEE standards} 

Despite the challenges, developing TEE standards seems too important to be ignored considering the stake behind them. Both the developer and the TEE research communities perhaps would agree that the future of TEEs will not be ideal if they remain to be vendor locked and fragmented. 

\section{Conclusion}\label{sec-conclusion}
TEEs have been proved a powerful tool for constructing secure applications. 
Using blockchain-based TEE applications as targets, we examine and summarize the key security models and assumptions of these applications. From this perspective, the paper provides an overall picture of research challenges and opportunities in this area. The lessons learned from these use cases provide some potential directions for future research and areas for improvement.

% , which allows the TEE researchers and designers to rethink some of the basic concepts in hardware TEE design. 


\bibliographystyle{ACM-Reference-Format}
\bibliography{references}

\end{document}